# Flash Colloidal Gold Nanoparticle Assembly in a Milli Flow System: Implications for Thermoplasmonic and for the Amplification of Optical Signals.

Florent Voisin,[1] Gérald Lelong,[2] Jean-Michel Guigner,[2] Thomas Bizien,[3] Jean-Maurice Mallet,[4] and Florent Carn[1*]

[1] Laboratoire Matière et Systèmes Complexes, UMR 7057, Université de Paris, CNRS, Paris, France.
[2] Institut de Minéralogie, de Physique des Matériaux et de Cosmochimie, UMR 7590, Sorbonne Université, CNRS, Paris, France.
[3] Synchrotron Soleil, Beamline SWING, Gif SurYvette, Cedex F-91192, France
[4] Laboratoire des Biomolécules, UMR 7203, École Normale Supérieure, PSL Research University, Sorbonne Universités, CNRS, Paris, France.

[*] E-mail : florent.carn@u-paris.fr

**Abstract**

The assembly and stabilization of a finite number of nanocrystals in contact in water could maximize the optical absorption per unit of material. Some local plasmonic properties exploited in applications, such as photothermia and optical signal amplification, would also be maximized which is important in the perspective of mass producing nanostructures at a lower cost. The main lock is that bringing charged particles in close contact requires the charges to be screened/suppressed, which leads to the rapid formation of micrometric aggregates. In this article, we show that aggregates containing less than 60 particles in contact can be obtained with a milli-flow system composed of turbulent mixers and flow reactors. This process allows to stop a fast non-equilibrium colloidal aggregation process at millisecond times after the initiation of the aggregation process which allows to control the aggregation number. As a case study, we considered the rapid mixing of citrate coated gold nanoparticles (NP) and $AlCl_3$ in water to initiate a fast aggregation controlled by diffusion. Injecting a solution of polycation using a second mixer allowed us to arrest the aggregation process after a reaction time ($t_Q$) by formation of 'overcharged' cationic aggregates. We obtained within seconds stable dispersions of a few milliliters composed of particle aggregates. Our main result is to show that it is possible to master the average aggregation number ($\overline{N}_{agg}$) between ~ 2 and ~ 60 NP/aggregate by varying $t_Q$ between ~ 10 ms and ~ 1 s. The relationship between $\overline{N}_{agg}$ and $t_Q$ is linear as expected for the so-called diffusion-limited cluster aggregation process. In particular, we were able to stabilize aggregates with $\overline{N}_{agg}$ ~ 10 NP/aggregate showing a strong plasmonic coupling giving rise to an optical absorption band whose maximum is located at 800 nm. Such aggregates of readily available compounds are interesting in the perspective of maximizing the optical absorption per unit of gold at the wavelength of commercial lasers used in the fields thermoplasmonic and surface enhanced Raman scattering.

# 1. Introduction

It has become common to conduct colloidal chemistry in a continuous milli or micro flow reactor.[1–4] This is often motivated by the assertion that such flow processes are safer, more sustainable and better suited to automation than batch processes.[5] Flow chemistry also has the advantage of allowing reactions that cannot be performed in batch, thus opening new possibilities of synthesis. This is notably the case of a particular field of flow chemistry called "flash" chemistry.[6,7]

This field has emerged over the last decade, at the instigation of J.I. Yoshida, for the synthesis of organic molecules on a preparative scale by fast reactions which cannot be achieved by batch chemistry.[6,7] The characteristic times of these reactions involving highly reactive, unstable, species are of the order of a second or less. The principle of the method is simple. The set-up consists of micromixers and micro flow reactors. The reactants are first mixed in a fast and turbulent manner in the micromixers. Then, the reaction proceeds as mixed reactants travel through the flow microreactor. The reaction time, noted $t_R$, can be precisely controlled, at the millisecond scale, by adjusting the length of the flow reactor and the speed of co-injection.[8]

In the field of colloidal science, fast micromixers and micro flow reactors have been used in combination with synchrotron techniques to probe the nucleation and growth of nanoparticles[9–13] or their assembly.[14–16] This type of flow process has also been used to better control the synthesis of nanoparticles with the so-called flash nanoprecipitation approach.[17] This consists of simultaneous incorporation of multiple species, comprising the reactants and the stabilizing agent, in a turbulent micromixer to achieve high super-saturation and nucleation rate for nano-particle formation in the millisecond range. These studies on colloidal systems were developed during the same period as the work initiated by Yoshida's team in the field of organic chemistry but, independently, without reference to it. Therefore, to our knowledge, the principle of associating different micro mixers separated by micro flow reactors to control fast colloidal assembly processes taking place out of equilibrium has never been investigated so far.

The purpose of the present work is to demonstrate the potential of such 'flash *colloidal* chemistry' by considering the assembly of citrate coated gold nanoparticles (AuNP) induced by the rapid addition of a multivalent electrolyte (i.e. $AlCl_3$) at a concentration well above the critical coagulation concentration.[18–20] The rapid mixing of these two partners in a first micromixer initiates an aggregation process controlled by diffusion[21–23] leading to the formation of micrometric fractal aggregates of nanoparticles after a few seconds. Injecting a polyelectrolyte (PEL) solution of quaternized chitosan (QC) using a second micromixer, located at a certain distance from the first, allowed us to arrest (i.e. quench) the aggregation process after a given reaction time $t_Q$ (Scheme 1b). By this process, we were able to obtain, in a single experiment, stable dispersions of a few milliliters composed of fractal aggregates.

Our main result is to show that it is possible to master the average aggregation number ($\overline{N}_{agg}$) between ~ 2 and ~ 60 NP/aggregate by varying $t_Q$ between ~ 10 ms and ~ 1 s. The relationship between $\overline{N}_{agg}$ and $t_Q$ is linear as expected for the so called diffusion limited cluster aggregation process (DLCA).[21,22] In addition, the structure and optical properties of the aggregates obtained at $t_Q \geq 4$ s are similar to those of the aggregates prepared by a 'manual' batch approach. These characteristics show that the rapid mixing process implemented in the flash assembly process does not affect the aggregation mechanism and allows the structure of the aggregates to be frozen at a given stage of their rapid growth. Thus, we were able to stabilize aggregates of about 10 particles showing a strong plasmonic coupling giving rise to an optical absorption band whose maximum is located at 800 nm.

From an applicative point of view, this particular result is interesting because numerical simulations have shown that the amplification of the optical absorption per unit of gold should be maximized at 800 nm for structures composed of chain fragments, containing about ten particles in close contact.[24] Beyond 10 particles per aggregate, the amplification factor reaches a plateau due to the so called 'infinite chain effect'.[25–27] Thus, the fact that small aggregates can be formed rapidly in water, without single particles in equilibrium, should allow to maximize the heating (i.e. thermoplasmonic) and detection (i.e. surface enhanced Raman scattering (SERS)) capabilities per unit of gold with current commercial lasers operating around 800 nm.

## 2. Materials and Methods

**2.1. Materials.** Gold (III) chloride trihydrate ($HAuCl_4 \cdot 3H_2O$, > 99.99 %), trisodium citrate dihydrate ($Na_3C_6H_5O_7 \cdot 2H_2O$, ≥ 99 %), chitosan ($C_8H_{13}NO_5$, degree of N-deacetylation ~ 75 %, $M_w$ ~ 310-375 kg/mol according to the supplier) were purchased from Sigma-Aldrich. Glycidyltrimethylammonium chloride, noted GTMAC ($C_6H_{14}ClNO$, 80 wt% in water) was purchased from Tokyo Chemical Industry. Aluminium chloride was purchased from Sigma-Aldrich ($AlCl_3$, > 99 %). All solutions were prepared with milliQ water (R = 18.2 MΩ). Details concerning the synthesis of quaternized chitosan are presented in supporting information (SI).

**2.2. Synthesis of citrate coated gold particles.** Citrate stabilized gold nanoparticles were first synthesized according to the seeded growth protocol proposed by Bastús et al.[28,29]. Briefly, 97 mg of citrates (0.33 mmol) were put in 150 mL of water (2.2 mM solution) and refluxed for 15min, before adding 1 mL of a 10 g/L solution of $HAuCl_4.3H_2O$. The solution was kept on reflux for 10 min, then the heating was slowly cooled down to 90 °C. The seeds had their size increased by repeating several growing cycles as followed: 55 mL of the solution were withdrawn, followed by the addition of 53 mL of water and 1 mL of a 60 mM citrates solution. As soon as the temperature reached 90 °C again, 1 mL of $HAuCl_4$ were added and the solution was stirred for 30 min. Then, 1 mL of $HAuCl_4$ were added and the solution was stirred again for 30 min. These growing cycles were repeated up to the desired size of AuNPs. Typically, 3 cycles were required to obtain particles with an hydrodynamic diameter $D_H \sim 39$ nm (D = 23.5 ± 2.2 nm in TEM). Between the seeds and the fourth cycle, theoretical gold nanoparticles concentration varies from $3.0 \times 10^{12}$ NPs/mL to $1.2 \times 10^{12}$ NPs/mL. These theoretical concentrations were calculated under the assumption that: all the gold present in the medium is included in the nanoparticles (i.e. 100% synthesis yield), the particles are spherical and gold colloid density is equal to gold in bulk. We underline that the concentration values deduced from the SAXS measurements agree well with the theoretically calculated values (Figure S1.e). The characterizations of the nanoparticles used in this study are presented in figure S1 of SI.

**2.3. Stop-flow and Quench flow experiments.** The 'stop-flow' (Scheme 1.a) and 'quench-flow' (Scheme 1.b) experiments were made at 25 °C with a SFM-4000 apparatus from BioLogic (Grenoble, France). The solutions (AuNP, Quaternized chitosan, $AlCl_3$) are contained in vertical syringes (V = 10 mL) filled from the top to allow the evacuation of the bubbles formed during the filling. The syringes are driven by independent stepping motors generating a flow from the bottom to the top. The fast mixing are made within a short time window (~ 10 ms) using turbulent mixers of type Berger-Ball.[30] In the 'quench-flow' configuration (Scheme 1.b) a syringe is placed after the sample cell, to collect the mixed sample at the end of the injection.

**2.4. Small Angle X-Ray Scattering (SAXS).** The SAXS experiments were performed on the SWING beamline at SOLEIL synchrotron (Saint-Aubin, France) with a photon energy of 12 keV and a sample-to-2D detector (Dectris Eiger 4M) distance of 6.3 m leading to the following q-ranges: $0.0016 \leq q$ (Å$^{-1}$) $\leq 0.23$. We recall that the norm of the scattering wave vector is $q = (4\pi/\lambda)\sin(\theta/2)$, where θ is the scattering angle and λ = 1.033 Å is the wave-length. All the intensities are expressed in differential cross-section per volume unit:

$$I(q) = \frac{dN(q)}{N_{Tot.}} \frac{1}{\Delta\Omega.\delta d}$$

where $\delta d$ is the sample' thickness, $dN(q)/N_{Tot.}$ is the probability for a transmitted photon to be scattered with a momentum transfer $q$ and collected at the solid angle $\Delta\Omega$. $dN(q)$ was measured after azimuthal integration of signal collected on the detector, and $N_{Tot.}$ was estimated from the beam transmitted to a photodiode mounted after the sample. The detector and photodiode were calibrated using water as a reference. The absolute water scattering intensity could be calculated as follows: $I = \rho^2 kT\chi \approx 1.63 \times 10^{-2} cm^{-1}$ at 293 K, with the scattering length density $\rho \approx 9.388 \times 10^{10} cm^{-2}$, the Boltzmann constant $k$, the temperature $T$ in $K$ and the isothermal compressibility $\chi \approx 10^{-10} Pa^{-1}$. The water contribution has been subtracted from all normalized scattering patterns. Measurements on quenched samples were done *ex situ* by inserting the samples into cylindrical quartz capillaries of 1.5 mm diameter that were sealed and left vertical in the field of gravity. Typically, 10 successive frames of 0.5 s each were recorded and compared to check the absence of beam damage. The scattering patterns were always isotropic. Each frame was first angularly averaged over all frames and the pure solvent spectrum was subtracted. For *in situ* stop-flow experiments, the SFM-4000 apparatus was connected at the output to a flow through quartz capillary (Hilgenberg GmbH, Germany) of 50 μm wall thickness and 2 mm inner diameter. The acquisition time per data frame, was set to 38 ms to avoid radiation damage to the sample.

**2.5. Cryogenic Transmission Electronic microscopy (cryo-TEM).** Images were taken on an Ultrascan 2 k x 2 k CCD camera (Gatan, USA), using a LaB 6 JEOL JEM 2100 (JEOL, Japan) cryo-microscope operating at 200 kV with a JEOL low dose system (Minimum Dose System) to protect the thin ice film from any irradiation before imaging and to reduce the irradiation during the image capture. The images were recorded at 93 K. The size of the aggregates was measured on the images manually using ImageJ software. Our procedure was to determine an average diameter by averaging the length of the aggregates in two perpendicular directions. Knowing that this measurement procedure in 2D on rather polydisperse fractal clusters could lead to potential errors/overestimations, we also systematically determined the number of aggregation that was readily available with minimal error due to the small size of the aggregates. We did not try to estimate the 2D fractal dimension by cryo-TEM knowing that we could obtain the mass fractal dimension at 3D by SAXS.

**2.6. Ultraviolet visible near-infrared absorption spectroscopy (UV-vis-NIR).** Measurements were performed on a Perkin Elmer Lambda 1050 spectrophotometer. 5 mm thickness Hellma cell (quartz) were filled in by the diluted AuNPs, QC and PEL. The diffuse transmission was measured by collecting the light exiting the sample using an integrating sphere (Perkin Elmer Accessory). Spectra were corrected from the quartz cell filled in with water.

## 3. Results and discussion
### 3.1. Multivalent electrolyte induced particle assembly in a turbulent flow

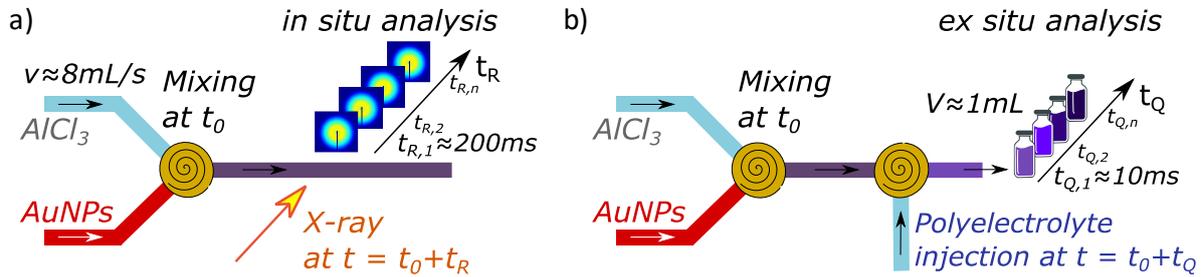

**Scheme 1.** Illustration of the fast milli-fluidic mixing configurations used in this work: (a) stop-flow configuration combined with in situ, and time resolved, SAXS measurements; and (b) quench-flow configuration.

The gold NPs assembly was obtained by mixing equal volumes (V = 400 µL) of two aqueous solutions containing citrated coated gold particles ($C_{NP}$ = 4.27×10$^{11}$ NP/mL) and an aluminium salt ($C_{AlCl_3}$ = 20 mM). The aluminum salt concentration is higher than the critical coagulation concentration[31] to obtain a rapid particle aggregation that should correspond to a fast diffusion-limited cluster aggregation (DLCA) process[21–23,18]. We recall that there are 2 peculiar regimes for colloid aggregation that are differentiated by particle-particle interactions. These two regimes are DLCA and reaction-limited cluster aggregation (RLCA). These regimes were traditionally called fast and slow aggregation respectively. In DLCA the diffusing particles move by Brownian motion without particle-particle interactions except when they come into contact. In this case, the contact leads systematically (i.e. probability of bonding equal to 1) to an irreversible bonding. The resulting aggregate continues diffusing and grow as soon as it comes into contact with a particle or other aggregate.

For this system, the aggregation should result from both a strong decrease of the Debye screening length and from the coordination of citrate (i.e. hard base) and Al$^{3+}$ (i.e. hard acid). The fast mixing was made by co-injection of the two solutions at ~ 8 mL/s in a turbulent mixer, of type Berger-Ball[30], using a SFM-4000 apparatus from BioLogic. The particle assembly was characterized with nanoscale and millisecond resolutions by synchrotron *in situ* SAXS on the SWING beamline at Soleil synchrotron (Scheme 1.a).[32] This method enabled us to study the nanostructure formation over several length scales (i.e. 3 - 300 nm) from 200 ms after mixing (Figure 1a). The intensity scattered by an assembly of spheres of volume V is sensitive to their average size through the sphere form factor P(q) and their organization through the average structure factor S(q) as follows:

$$I(q) = \Phi.V.(\Delta\rho)^2.P(q).S(q) \quad (1)$$

where $\Phi$ is the volume fraction of the particles, q is the norm of the scattering vector and $\Delta\rho = \rho_{Au} - \rho_{eau}$ is the contrast of X-ray scattering length density ($\rho$) between gold and water.

At the shortest observed reaction time (i.e. 200 ms), the scattering curve is close to the form factor of non-interacting (i.e. $S(q) = 1$) polydisperse gold spheres with a mean radius $\overline{R}$ = 12.7 nm and a log. normal size distribution (polydispersity: σ = 0.1). As the aggregation proceeds, $I(q)$ increases with time for $q < 0.01$ Å$^{-1}$ whereas all the curves remain superimposed for q ≥ 0.02 Å$^{-1}$ (i.e. $q\overline{R} \gg 1$). The latter behavior shows that the particle' concentration and characteristics remain the same during the studied period. The increase of $I(q)$ with time in the low q regime is due to the aggregation process (i.e. $S(q) \neq 1$). In order to estimate the average structure factor $S(q)$ using equation (1), the scattering patterns were divided by the simulated form factor of non-interacting polydisperse gold spheres at the studied concentration. We emphasize that we chose to use a function obtained by simulation rather than the I(q) function obtained experimentally by turbulent mixing of water and particles (Figure S1.e) because the latter curve shows an upturn at small q (q < 0.002) due to subtraction. This kind of behavior is commonly found in other curves obtained at different times. As shown in Fig. 1b, this procedure yields $S(q,t)$ with the assumption that the virial effects between aggregates are negligible at the considered concentration that is quite low. Overall, the curves present a similar profile with a Guinier plateau at low q for short time measurements and a power law decay toward unity at higher q values any time after mixing. The aggregation dynamic is characterized by the increase of $S(q \to 0)$ and of the power law exponent which reflect the increase of the averaged aggregation number ($\overline{N}_{agg}$) and of the aggregate mass fractal dimension (d$_f$) respectively. We recall that the mass fractal dimension characterizes the spatial distribution of the mass within the aggregate. For particles with a mass, M$_0$, and a size, R$_0$, the total mass of the assembly, M, scales with its radius, R, measured from any site within the fractal structure, as $M(R) \propto M_0 \left(R/R_0\right)^{d_f}$. d$_f$ ~ 2 is typical of a random branched structure while d$_f$ ~ 1 and d$_f$ ~ 2.9 are expected for 1D and 3D structures respectively.[33,34]

Since the Guinier regime is not observed for all times, we have characterized the dynamic of aggregation by plotting the evolution of $S$(q = 0.002 Å$^{-1}$, t) in Figure 1c. As expected for diffusion-limited cluster aggregation, $S(q = 0.002$ Å$^{-1}, t)$ scales as a power law over time but with an exponent ~ 0.6 whereas $\overline{N}_{agg}$ would normally scale as $\propto t$ for such aggregation process. Indeed, as shown by Weitz et al.[35], the DLCA growth rate for a mean cluster mass, $\frac{dM}{dt}$, is determined by the diffusion equation: $dt \sim (8\pi RCD)^{-1}$ where $C$ is the particle concentration,

$R$ the mean cluster radius and $D$ the translational diffusion coefficient. Assuming that, in a mean-field treatment of the dynamics, the dominant growth occurs by aggregation of two clusters of roughly equal size, so that $dM \sim M$, the authors showed that $R \propto t^{1/d_f}$ for $R \gg R_0$. As a consequence, one derive the following relations: $M \propto \overline{N}_{agg} \propto R^{d_f} \propto t$.

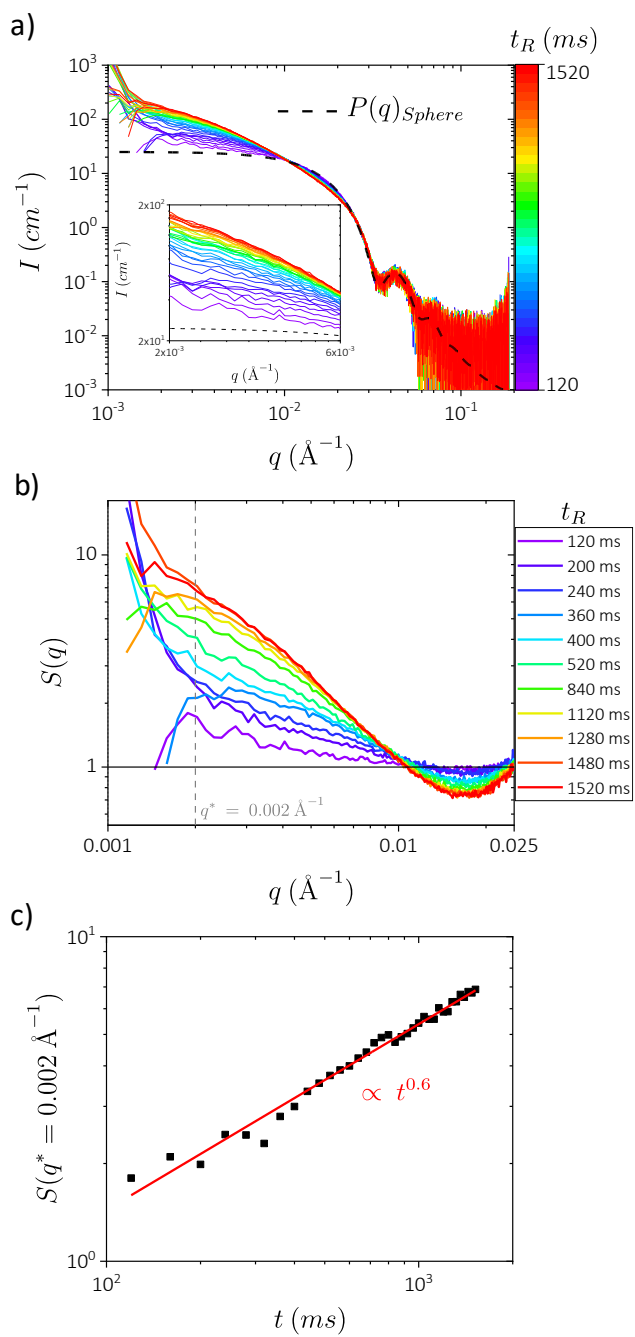

**Figure 1.** a) Time evolution of the X-ray scattering patterns (I(q)) obtained *in situ* during gold NPs assembly between 120 and 1520 ms after fast mixing with aluminium salt ($C_{AlCl3}$ = 20 mM). The black dashed line corresponds to the form factor of sphere (P(q)) with a log normal size distribution (R = 12.7 nm; PDI = 0.1). b) Time evolution of the structure factor (S(q)) derived from some of the curves shown in figure1a by dividing I(q) with P(q). c) Time evolution of the structure factor shown in figure1b at q = 0.002 Å$^{-1}$.

In the same way, the mass fractal dimension of the aggregates tends towards 1.2 whereas a value of 1.8 is expected for DLCA. These slight differences may reflect the contribution of hydrodynamic stress related to the turbulent flow in the micromixer[36]. We emphasize that the process of particle assembly has been repeated several times under the same conditions with little variations in the results (Figure S3).

**3.2. Stabilisation of colloidal oligomers at different stages of growth**

After the preliminary study concerning the assembly process, we tried to arrest the aggregation after different reaction times ($t_R$) by injecting a polyelectrolyte solution using a second micromixer (Scheme 1.b). The stabilization mechanism should be essentially of electrostatic origin. Indeed, the initial mixing of an aluminum salt with the negatively charged particles induces a rapid aggregation process where most collisions lead to irreversible sticking. As shown by electrophoretic mobility measurements (Figure S12), the growing aggregates which are globally neutral become positively charged following the addition of chitosan cationic chains. The stabilization must result from the formation of these 'overcharged' aggregates[37] with an overall positive charge sufficient to stop the aggregation process despite a high concentration of ions in the medium.

In practice, the 'quenching' step consisted of co-injecting 400 µL of a quaternized chitosan solution (C = 0.5 g/L) and 800 µL of the reaction medium containing the gold particles and the aluminum salt at ~ 8 mL/s. The time at which the quenching is made will be noted $t_Q$ in the following. The quenched samples ($V_{Tot.} \approx 1.2$ mL) were then collected in a syringe and analyzed afterwards (i.e. *ex situ* measurements) by SAXS, UV-Vis spectroscopy and cryo-TEM.

At first sight, the temporal evolution of the SAXS patterns obtained by arresting the aggregation process (Figure 2a) is comparable to the non-perturbed *in situ* dynamic (Figure 1a). Overall, this evolution corresponds to the growth of fractal structures with constant particle concentration (i.e. no sedimentation or phase separation). One specificity is that the fractal dimensions are globally higher with a smaller range of variation from 1 to 1.73 when $t_Q$ varies from 10 ms to 1s. At long time, the curve obtained at $t_Q \geq 1$s is superimposable to the curves obtained by batch preparation (i.e. injection and mixing using micropipettes and vortex mixer respectively) with or without quenching. It means that, at least for long times, the structures formed do not depend on the mixing mode nor the quenching agent. The increase of $d_f$ may therefore indicate that the quenched structures observed several minutes after formation have time to relax to an equilibrium state. Hence, contrarily to *in situ* characterization shown in figure

1, the value obtained at long $t_Q$ ($d_f$ = 1.73) corresponds to that expected for a DLCA-type process.

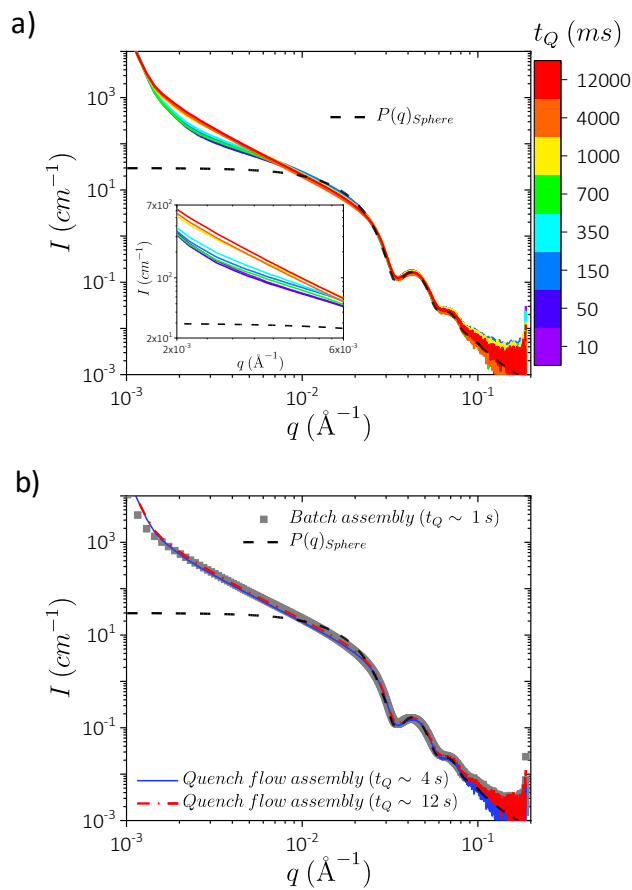

**Figure 2.** a) X-ray scattering patterns (I(q)) of gold NPs assemblies obtained by quenching the fractal growth at different times ($t_Q$) indicated on the plot. The NP aggregation has been initiated by fast mixing with aluminium salt ($C_{AlCl_3}$ = 20 mM). All samples were characterized 5 min after mixing. b) X-ray scattering patterns of gold NPs assemblies obtained by two different procedures: quench flow and batch assembly. The quench flow corresponds to the protocol associated to figure 1a. The batch assembly corresponds to a protocol where aggregation is induced and stopped by manual mixing and injection with a micropipette for about 1 s after the initiation of aggregation. In the two figures, the black dashed line corresponds to the form factor of sphere (P(q)) with a log-normal distribution (R = 12.7 nm; PDI = 0.1).

We completed this structural study by using cryogenic transmission electron microscopy (cryo-TEM). As shown in figure 3 and figures S4-11 of S.I., the aggregates present rather polydisperse fractal structures whose average size visibly increases with $t_Q$ whereas the number of individual particles evolves in the opposite direction.

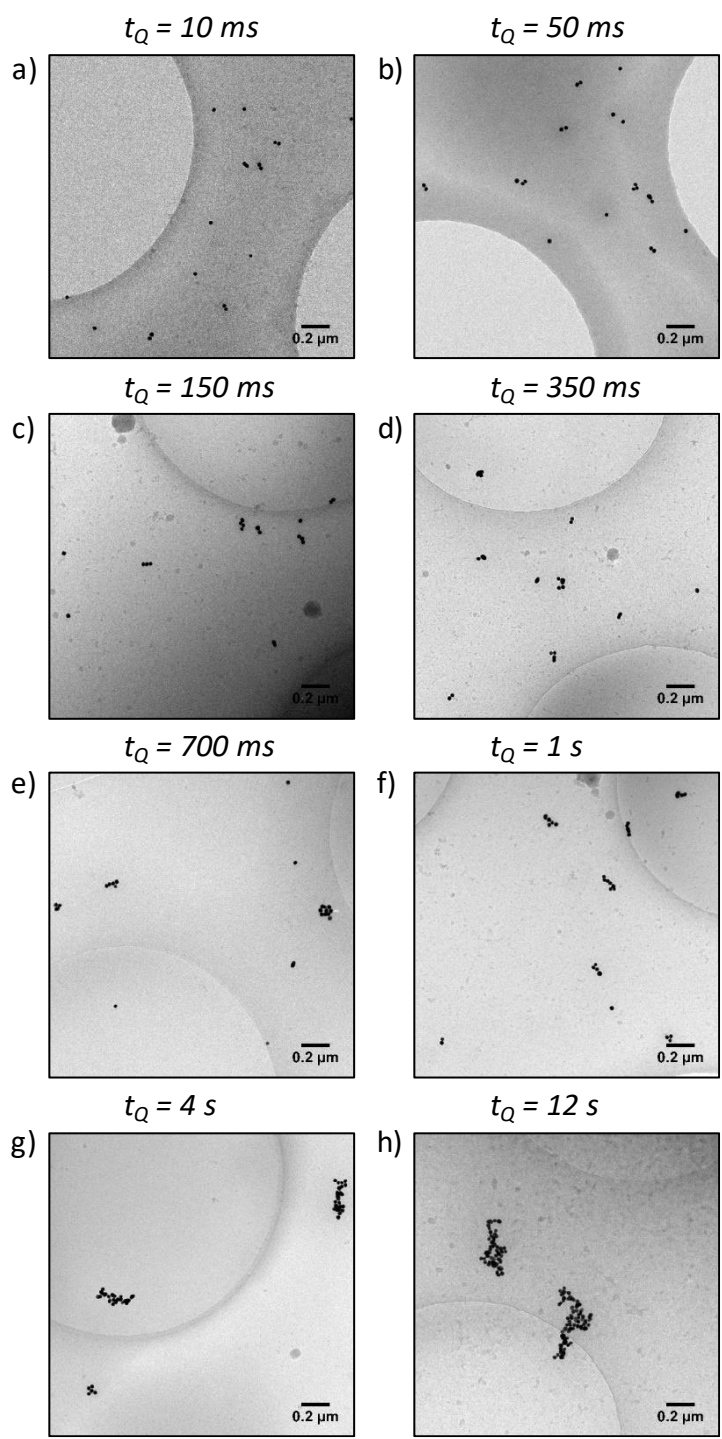

**Figure 3.** Typical cryo-TEM images of aggregates stabilized after different reaction times: (a) $t_Q = 10$ ms, (b) $t_Q = 50$ ms, (c) $t_Q = 150$ ms, (d) $t_Q = 350$ ms, (e) $t_Q = 700$ ms, (f) $t_Q = 1$ s, (g) $t_Q = 4$ s, (h) $t_Q = 12$ s. Additional images for each condition are provided in the supplementary information (Figures S4-S11). The samples were prepared 24 h before observation.

This is confirmed by a statistical analysis of the images (Figure 4a) showing a continuous variation of the average aggregate size ($\overline{D}$) from 40 to 300 nm when $t_Q$ varies from 10 ms to 12 s.

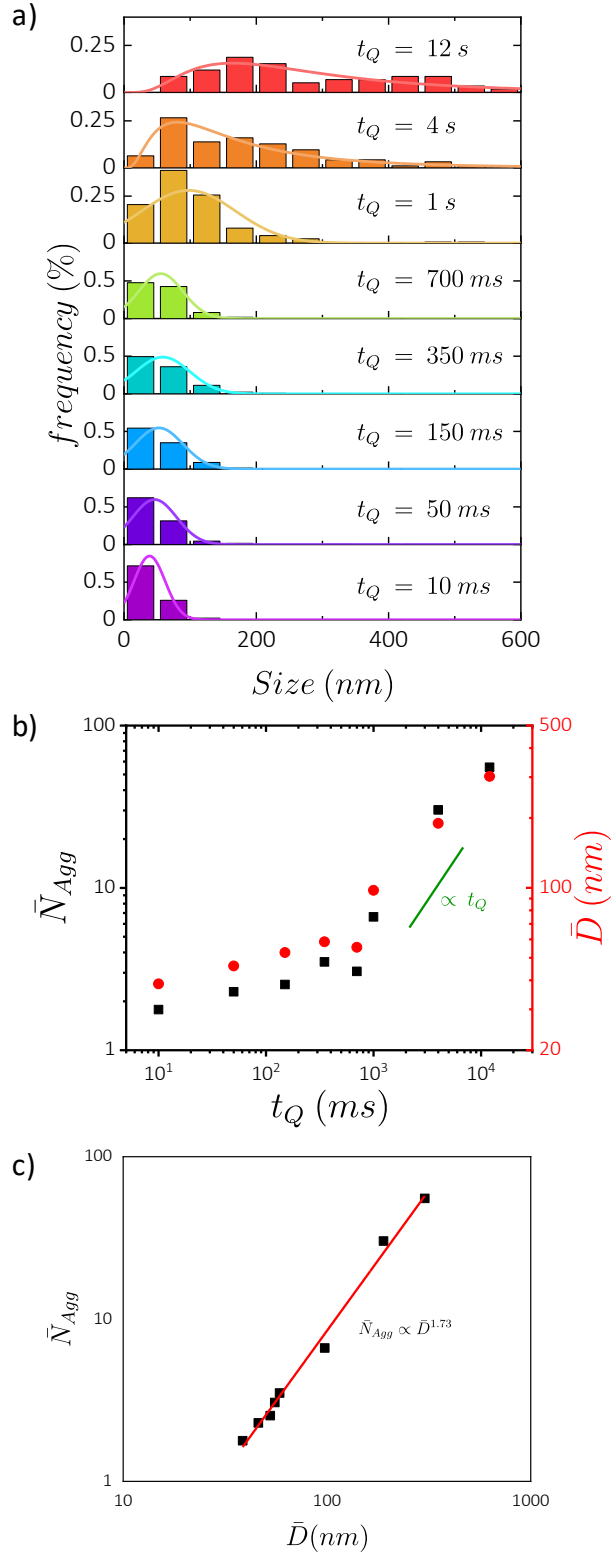

**Figure 4.** Statistical analysis of the sizes and aggregation numbers of the assemblies observed by cryo-TEM (Figure 3, S4-S11). a) Size histogram of aggregates stabilized after different quenching times ($t_Q$) indicated on the images. b) Average sizes ($\bar{D}$) and aggregation numbers ($\bar{N}_{Agg}$) as a function of $t_Q$. c) Average aggregation numbers versus average sizes.

This evolution is initially slow and then faster with a kinetic ($\overline{N}_{agg} \propto t_Q$ with $t_Q \sim t_R$) in line with those expected for DLCA (Figure 4b). From a structural point of view (Figure 4b), we also obtain a result consistent with a DLCA process since the aggregation number evolves in a power law with the number averaged size $\overline{D}$ such as: $\overline{N}_{agg} \propto \overline{D}^{1.73}$.

In summary, we show that the turbulent injection of polyelectrolyte chains allows to arrest a fast DLCA type aggregation process by 'overcharging' the growing aggregates. This approach allows to obtain gold particle assemblies whose average aggregation number is controlled between ~ 2 and ~ 60 NP/aggregate by the quenching time. The evolution of the averaged characteristics of the quenched aggregates (*i.e.* $\overline{N}_{agg}$ and $\overline{D}$) with the quenching time is in agreement with the results expected for a DLCA type dynamic. Moreover, the structure and optical properties of the aggregates obtained at $t_Q \geq 4$ s are similar to those of the aggregates prepared by a 'manual' batch approach. Overall, these results show that the rapid mixing process implemented in the flash assembly process does not affect the aggregation mechanism and allows the structure of the aggregates to be frozen at a given stage of their rapid growth.

### 3.3. Optical properties of the colloidal oligomers

We monitored the evolution of the optical properties of colloidal oligomer dispersions as a function of $t_Q$ by means of UV/Vis/NIR spectroscopy (Figure 5). The measurements were made on the same set of samples shown in figures 2, 3 and S4-S11. The dynamics were characterized by a decrease of the intensity corresponding to the localized surface plasmon resonance (LSPR) of isolated particles at 524 nm and the appearance of a new plasmon band which gradually increases in intensity. One can also note a drop in extinction at λ = 400 nm (i.e. $t_Q$ = 12 s) that should be related to the sedimentation of large aggregates composed of a dense material. Indeed, the acquisition time of UV-Vis.-NIR spectra was about 300 s while the sedimentation speed of aggregates can be estimated to be higher than 1 μm/s for aggregates of size higher than 300 nm. Moreover, despite the subtraction of the spectra by the signal of a cell filled with pure milliQ water, all spectrum show a significant background (i.e. the extinction is non zero at high wavelength). This background can have different origins:

- Our integration sphere only collect the light scattered after transmission. Thus, we do not collect the backscattered intensity.

- The scattering increases with the size of the aggregates. This increase is less strong at 800 nm than for smaller wavelengths but the effect is probably not negligible.

- The citrates molecules may contribute to the background.

Interestingly, the second plasmon band is observed over a wavelength range from 600 nm to 950 nm when $t_Q$ varies from 10 ms to 12 s. This evolution is consistent with cryo-TEM observations showing a progressive formation of short linear aggregate which connect over time into ramified structures (Figure 3, S4-S11). We underline that this structuration leads to an efficient coupling of the dipolar modes in comparison with experiments[38] and numeric simulations[39] performed for comparable fractal aggregates of similar particles. A remarkable point is that the wavelength associated with the second extinction maximum rarely exceeds 700 nm for particles of this size assembled in this way, whereas in our case the second maximum is around 800 nm for $t_Q \geq 1\ s$. This specificity is certainly related to the fact that the aggregation mediated by $Al^{3+}$ allows the inter-particle distance to be reduced to such an extent that in some cases one can detect coalescence events between adjacent particles whatever $t_Q$ (Figure 5b). We underline that we have characterized extensively the initial particle batches without detecting fused particles and that the use of a cryo-TEM to make the observations limits the risks of fusion induced by the electron beam.

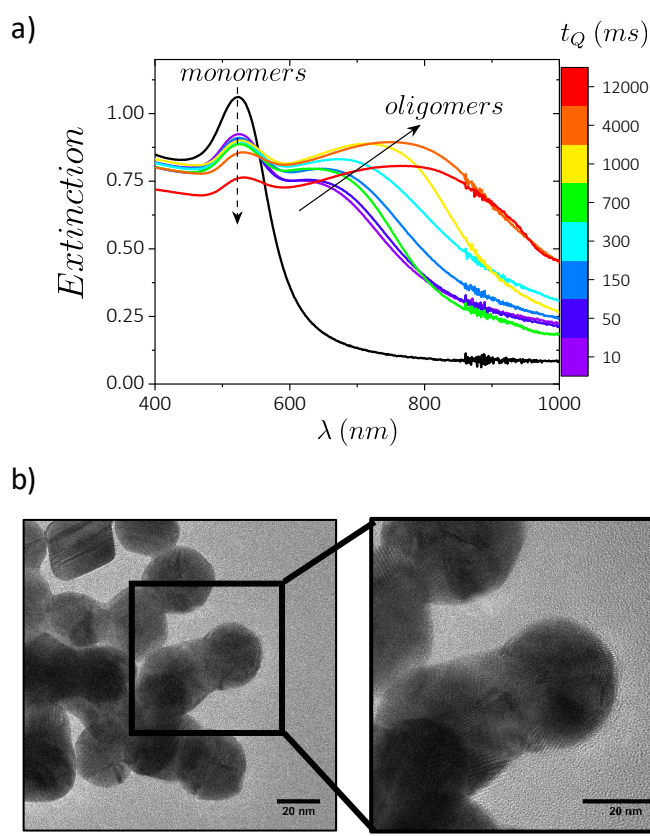

**Figure 5.** a) Extinction spectra of aqueous dispersions corresponding to the gold particles assemblies characterized in figure 2-4. b) High magnification TEM images of aggregates stabilized 12 s after the aggregation initiation.

From an applicative point of view, a remarkable result of this part concerns the possibility of stabilizing aggregates of about 10 particles (i.e. $t_Q \sim 1\ s$) showing an optical absorption band whose maximum is located at 800 nm. Indeed, according to numerical simulations the interest of this type of structure is to allow maximizing the amplification of the optical absorption per unit of gold.[24] Beyond ~10 particles per aggregate, the amplification factor reaches a plateau due to the so called 'infinite chain effect'.[25–27] Moreover, as shown in figure 3b and S9, at this stage of the aggregation the particles are almost all involved in aggregates which allows to fully exploit the plasmonic properties of all the particles present in the media.

In this context, the proposed approach of 'flash colloidal assembly' may be of interest for several applications in the fields of thermoplasmonic[40] and detection by surface enhanced Raman scattering (SERS)[41] where it is important to maximize the properties of the most common gold particles (i.e. Turkevich synthesis) to allow a massive use of these techniques minimizing both the cost, the amount of particles potentially released into the environment and the non-reproducible character.

Last but not least, the ramified structure of the aggregates formed is such that the surface of the particles is accessible to small molecules, which is an advantage for photocatalytic reactions[42] essential for sustainable technologies.

## 4. Conclusion

In this work, we show that a milli-flow system composed of turbulent mixers and flow reactors could enable to control a fast colloidal aggregation process taking place out-of-equilibrium. This so-called *fast colloidal assembly* approach extends the concept of flash chemistry developed in the last decade for the synthesis of organic molecules by fast reactions.

As a case study, we considered the assembly of citrate coated gold nanoparticles (NP) induced by the rapid addition of a multivalent electrolyte (i.e. AlCl$_3$). The rapid mixing of the partners initiates a fast aggregation controlled by diffusion. Injecting a solution of polycation (i.e. quaternized chitosan) using a second mixer allowed us to arrest the aggregation process after a given reaction time ($t_Q$) by formation of 'overcharged' cationic aggregates. By this process, we were able to obtain, at once, stable dispersions of a few milliliters composed of particle aggregates. Our main result is to show that it is possible to master the average aggregation number ($\overline{N}_{agg}$) between ~ 2 and ~ 60 NP/aggregate by varying $t_Q$ between ~ 10 ms and ~ 1 s. The relationship between $\overline{N}_{agg}$ and $t_Q$ is linear as expected for the so-called diffusion-limited cluster aggregation process. In addition, the structure and optical properties of the aggregates obtained at $t_Q \geq 4$ s are similar to those of the aggregates prepared by a 'manual' batch approach. These characteristics show that the rapid mixing process allows the structure of the aggregates to be frozen at a given stage of their rapid growth. In particular, we were able to stabilize aggregates of about 10 particles showing a strong plasmonic coupling giving rise to an optical absorption band whose maximum is located at 800 nm. The fast formation of this type of structure in water with common compounds is important in the perspective of maximizing the optical absorption per unit of gold at the wavelength of commercial lasers used in the fields thermoplasmonic and surface enhanced Raman scattering (SERS).

Overall, we believe that our results provide new opportunities to direct fast out-of-equilibrium aggregation by allowing the formation of finite size assemblies in which the inter-particle distance is minimal and the surface chemistry is adjustable.

## Acknowledgements.

ANR (Agence Nationale de la Recherche) and CGI (Commissariat à l'Investissement d'Avenir) are gratefully acknowledged for their financial support of this work through Labex SEAM (Science and Engineering for Advanced Materials and devices). F.C. and F.V. thank the ANR Coligomere-18-CE0006 for funding. We are indebted to N. Sanson (Soft Matter Sciences and Engineering, ESPCI, PSL University, Sorbonne Université, CNRS) for his assistance during preliminary TOC measurements. We acknowledge the technopolym platform (Université de Toulouse III-Paul Sabatier) for SEC and TOC measurements. We acknowledge the ImagoSeine core facility of the Institut Jacques Monod, member of the France. BioImaging infrastructure (ANR-10-INBS-04) and GIS-IBiSAWe also acknowledge synchrotron SOLEIL (SWING beam line, Saint-Aubin, France) for the SAXS beam time allocation. Last but not least, we acknowledge J. Mangue (Laboratoire BioLogic, Seyssinet-Pariset, France) for its welcome and its advice concerning the use of the SFM-4000 apparatus.

## Supporting information.

The supporting information concerns: (1) Synthesis and characterization gold nanoparticles and quaternized chitosan; (2) Study of aggregation kinetics by time-resolved SAXS measurements in stop-flow configuration; (3) Cryo-TEM observations of the aggregates obtained at different quench times; and (4) electrophoretic mobility measurements before and after the quenching of the aggregation by QC addition.

# Graphic for manuscript

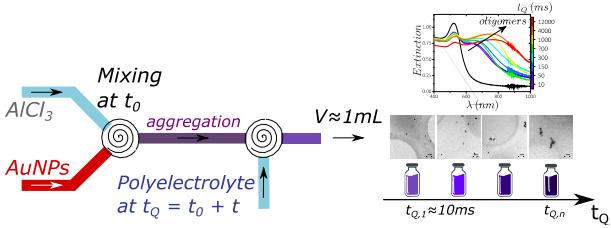